\documentclass[fleqn,twoside]{article}
\usepackage{amssymb,amsfonts,espcrc2}
\usepackage{graphicx}

\newcommand{\ind}[2]{^{#1}_{\mbox{\scriptsize #2}}}
\newcommand{\al}[2]{\alpha\ind{#1}{#2}}
\newcommand{\ASL}[2]{A\ind{#1}{{\sc sl,}{\tiny\,}#2}}
\newcommand{\ATL}[2]{A\ind{#1}{{\sc tl,}{\tiny\,}#2}}

\def\KL{K\"all\'en--Lehmann }
\def\Nc{N_{\mbox{\scriptsize c}}}
\def\nf{n_{\mbox{\scriptsize f}}}
\def\mpi{m_{\pi}}

\title{Impact of the pion mass on nonpower expansion for QCD observables}

\author{A.V.~Nesterenko\address[UV]{Departamento de F\'\i sica Te\'orica
and IFIC, Centro Mixto, Universidad de Valencia--CSIC, \\
~$\,$E-46100, Burjassot, Valencia, Spain}\address[JINR]{Bogoliubov Laboratory
of Theoretical Physics, JINR, Dubna, 141980, Russian Federation}
and
J.~Papavassiliou\addressmark[UV]}

\begin{document}

\begin{abstract}
A new set of functions, which form a basis of the massive nonpower
expansion for physical observables, is presented in the framework of
the analytic approach to QCD at the four-loop level. The effects due
to the $\pi$~meson mass are taken into account by employing the
dispersion relation for the Adler function. The nonvanishing pion
mass substantially modifies the functional expansion at low energies.
Specifically, the spacelike functions are affected by the mass of the
$\pi$~meson in the infrared domain below few GeV, whereas the
timelike functions acquire characteristic plateaulike behavior below
the two--pion threshold. At the same time, all the appealing features
of the massless nonpower expansion persist in the considered case of
the nonvanishing pion mass.
\vskip-1mm
\end{abstract}

\maketitle

     The renormalization group (RG) method plays a key role in the
framework of the Quantum Field Theory (QFT) and its applications.
Indeed, one is able to handle reliably the strong interaction
processes at high energies by employing this method together with
perturbative calculations. However, such perturbative solutions to
the RG equation possess unphysical singularities in the infrared
domain, a fact that contradicts the general principles of the local
QFT, and significantly complicates the theoretical description and
interpretation of the intermediate- and low-energy experimental data.
Nevertheless, an effective way to overcome these difficulties is to
complement the perturbative results with a proper nonperturbative
insight into the infrared hadron dynamics.

     One of the sources of the nonperturbative information about the
strong interaction processes is the dispersion relations. The idea of
employing the latter together with perturbation theory forms the
underlying concept of the so-called analytic approach to QFT, which
was first proposed in the framework of Quantum
Electrodynamics~\cite{RedBLS}. Recently, this approach has been
extended to Quantum Chromodynamics (QCD)~\cite{ShSol}  and applied to
the ``analytization'' of the perturbative power series for the QCD
observables~\cite{APT1,APT2,DV04}. The term analytization means the
restoring of the correct analytic properties in the kinematic
variable of a quantity under consideration by making use of the \KL
integral representation (positive $q^2$ corresponds to a spacelike 
momentum transfer hereinafter)
\begin{equation}
\label{DefAn}
\Bigl\{F(q^2)\Bigr\}_{\mbox{$\!$\scriptsize an}} =
\int_{0}^{\infty} \!\frac{\varrho(\sigma)}{\sigma+q^2}\,
d \sigma
\end{equation}
with the spectral function defined by the initial (perturbative)
expression for the quantity at hand:
\begin{equation}
\varrho(\sigma) = \frac{1}{2\pi i}\, \lim_{\varepsilon \to 0_{+}}
\Bigl[F(-\sigma-i\varepsilon) - F(-\sigma+i\varepsilon)\Bigr].
\end{equation}
However, there are several ways to embody the analyticity requirement
into the RG formalism, that eventually has given rise  to different
models for the analytic running coupling.

     Thus, in the original model due to Shirkov and
Solovtsov~\cite{ShSol} the analyticity requirement~(\ref{DefAn})  is
imposed on the perturbative running coupling itself. At the one-loop
level this leads to
\begin{equation}
\label{ARCSS}
\al{(1)}{ss}(q^2) = \frac{4 \pi}{\beta_{0}}\!
\left(\frac{1}{\ln z} + \frac{1}{1-z}\right)\!,
\qquad z=\frac{q^2}{\Lambda^2},
\end{equation}
whereas at the higher loop levels the integral
representation of the \KL type
\begin{equation}
\label{ARCKL}
\al{}{}(q^2) = \frac{4 \pi}{\beta_{0}}
\int_{0}^{\infty} \frac{\rho(\sigma)}{\sigma+q^2}\, d \sigma
\end{equation}
holds for this invariant charge. Ultimately, the
prescription~\cite{ShSol} results in the infrared finite limiting
value for the running coupling (see papers \cite{APT1,APT2,DV04} and
references therein for the details).

     Another way to incorporate the analyticity condition into the RG
formalism is to impose the analyticity requirement~(\ref{DefAn}) on
the perturbative approximation of the $\beta$~function with
subsequent solution of the corresponding RG equation~\cite{AIC}. At
the one-loop level this leads to
\begin{equation}
\label{AIC}
\al{(1)}{an}(q^2) = \frac{4 \pi}{\beta_{0}} \frac{z-1}{z\,\ln z},
\end{equation}
whereas at the higher loop levels the running coupling at hand can be
represented in the form of the \KL integral~(\ref{ARCKL}) as well.
Here the invariant charge possesses the infrared enhancement, which
plays an essential role in applications of this model to the study of
the quark confinement~\cite{AIC} and the chiral symmetry
breaking~\cite{CSBAIC}. It is of particular interest to mention that
the explicit one-loop form of the analytic running
coupling~(\ref{AIC}) has recently been rediscovered~\cite{Schrempp},
proceeding from entirely different motivations.

     The dispersion relations also play an important role for the
congruous description of the hadron dynamics in the spacelike and
timelike regions. In particular, it has been argued that the
dispersion relation for the Adler function~\cite{Adler}
\begin{equation}
\label{AdlerDisp}
D(q^2,\mpi^2) = q^2 \int_{4\mpi^2}^{\infty}
\frac{R(s)}{(s + q^2)^2}\, d s
\end{equation}
provides a firm ground for comparing the perturbative results for
$D(q^2,\mpi^2)$ with the measurable ratio $R(s)$ of the
$e^{+}e^{-}$~annihilation into hadrons. Indeed, one can continue an
explicit expression for the Adler function into the timelike domain
by making use of the inverse relation
\begin{equation}
\label{AdlerInv}
R(s,\mpi^2) = \frac{1}{2 \pi i} \lim_{\varepsilon \to 0_{+}}
\int_{s + i \varepsilon}^{s - i \varepsilon}\!
D(-\zeta,\mpi^2)\, \frac{d \zeta}{\zeta},
\end{equation}
where the integration contour lies in the region of the analyticity
of the integrand (see Refs.~\cite{Rad82,KP82}).

     The only information about the Adler function~(\ref{AdlerDisp})
available from the perturbation theory is its behavior in the
asymptotical ultraviolet region $q^2 \to \infty$. Specifically, at
the $\ell$-loop level
\begin{equation}
\label{AdlerPert}
D(q^2) \simeq 1 + \sum\nolimits_{j=1}^{\ell}d_{j}\!
\left[a\ind{(\ell)}{s}(q^2)\right]^{j}\!,
\end{equation}
where the overall factor $\Nc\sum_f Q_{f}^{2}$ is omitted throughout,
$a\ind{}{s}(q^2)=\al{}{s}(q^2)\beta_{0}/(4\pi)$ is the perturbative
``couplant'', $\beta_{0} = 11 - 2 \nf /3$, $\nf$ is the number of
active flavors, and $d_1=4/\beta_{0}$, $d_2 \simeq
(4/\beta_{0})^{2}\, (1.99 - 0.12\,\nf)$, see
Refs.~\cite{GKL91,SurSa91} for the details. At the same time, the
dispersion relation~(\ref{AdlerDisp}) implies that $D(q^2,\mpi^2)$ is
the analytic function in the complex $q^2$-plane with the only cut
along the negative semiaxis of real~$q^2$ beginning at the two--pion
threshold. Thus, the perturbative approximation~(\ref{AdlerPert})
violates this condition due to unphysical singularities
of~$\al{}{s}(q^2)$. Nonetheless, this disagreement can be avoided
within the analytic approach to~QCD.

\begin{figure*}[ht]
\noindent\centerline{
\begin{tabular}{lr}
\parbox{72.5mm}{\includegraphics[width=70mm]{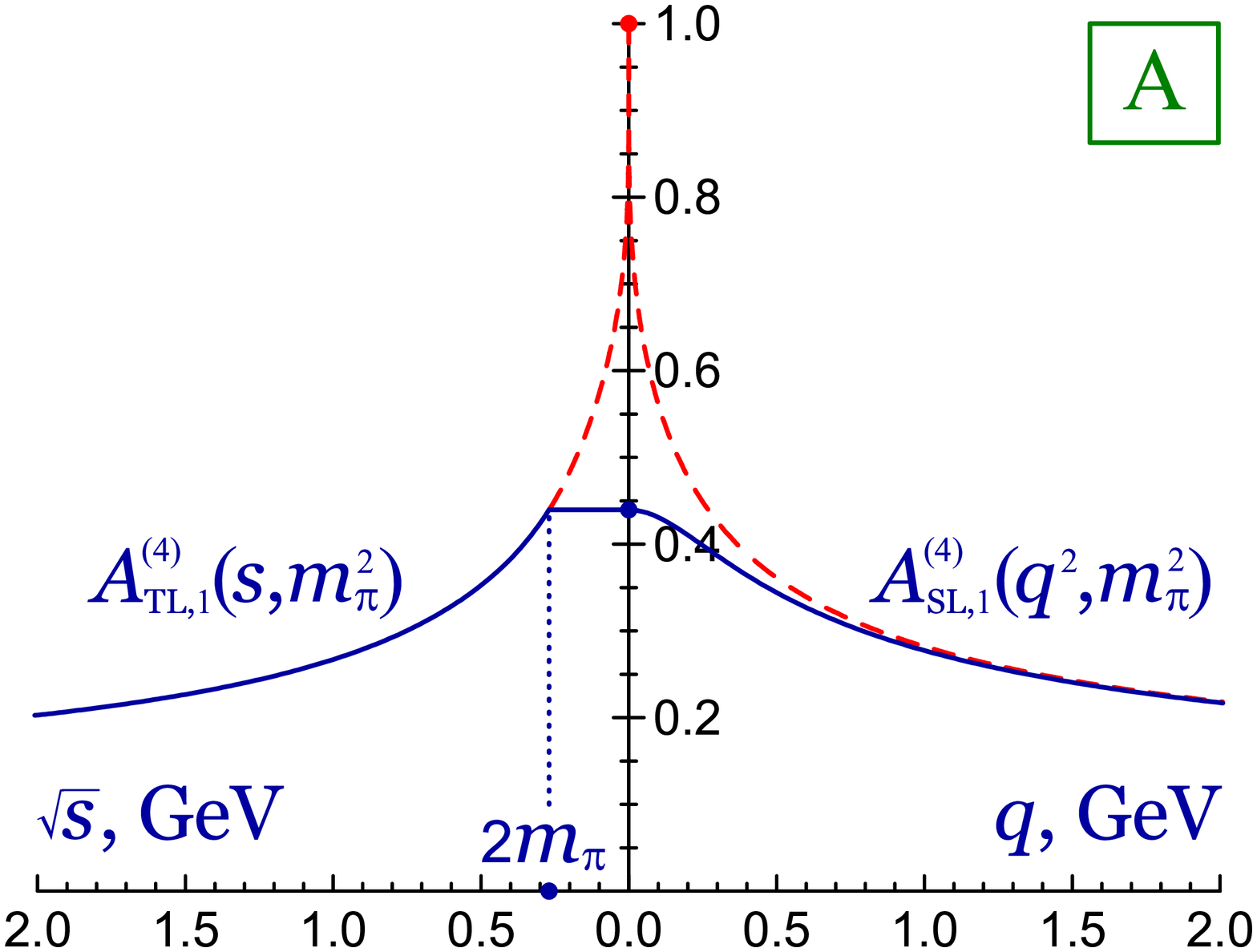}\hfill\vskip7mm}&
\parbox{72.5mm}{\hfill\includegraphics[width=70mm]{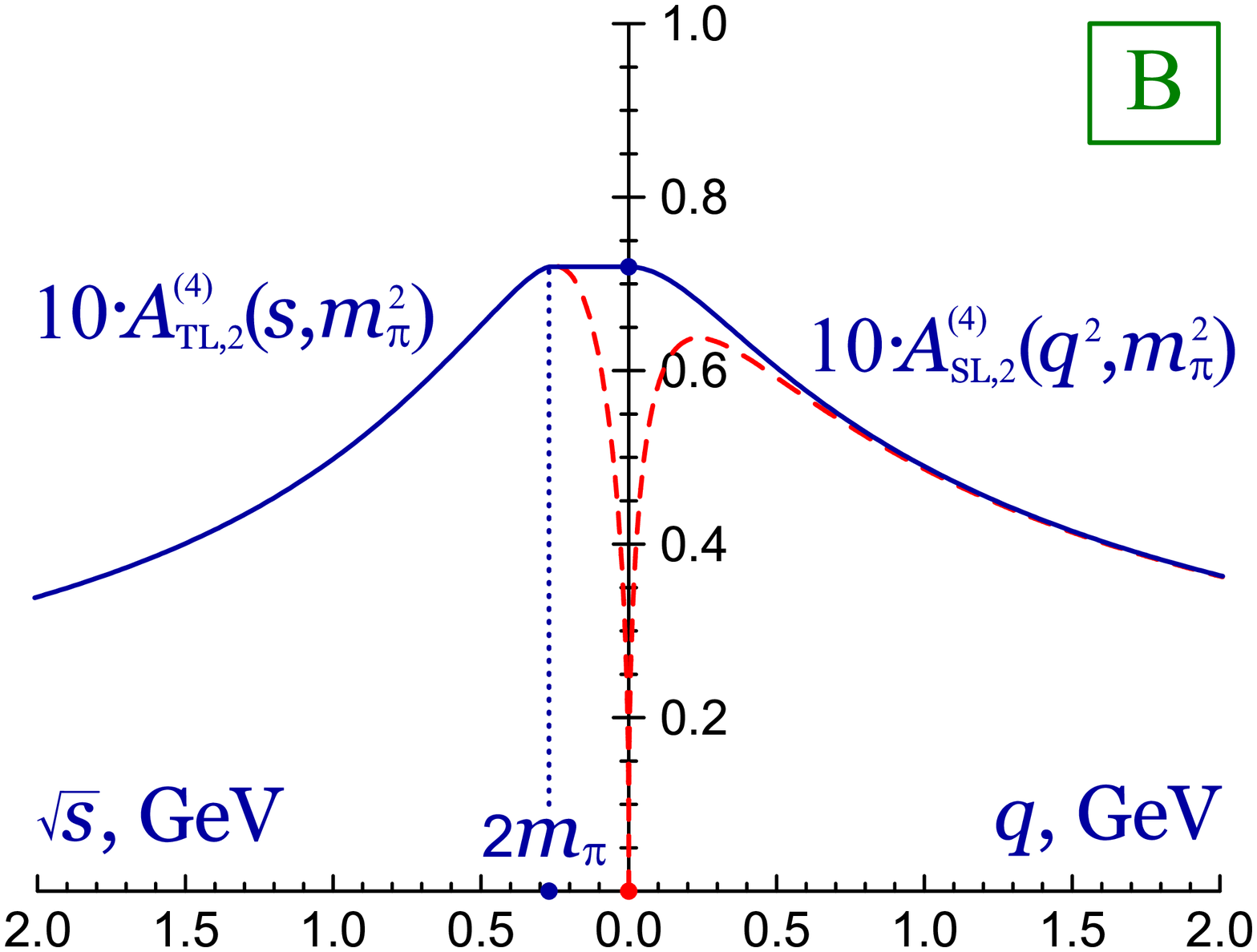}\vskip7mm}\\
\parbox{72.5mm}{\includegraphics[width=70mm]{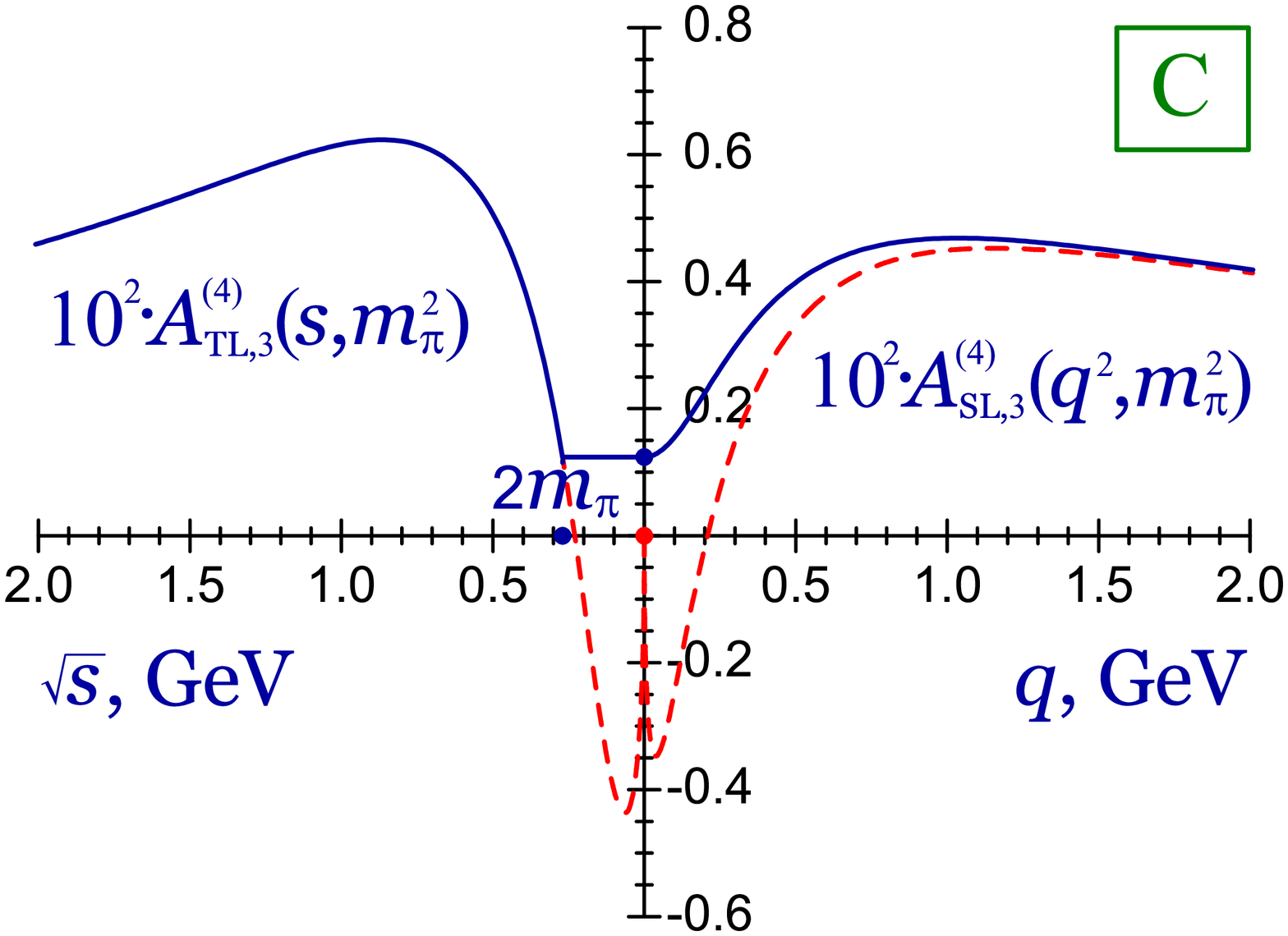}\hfill}&
\parbox{72.5mm}{\hfill\includegraphics[width=70mm]{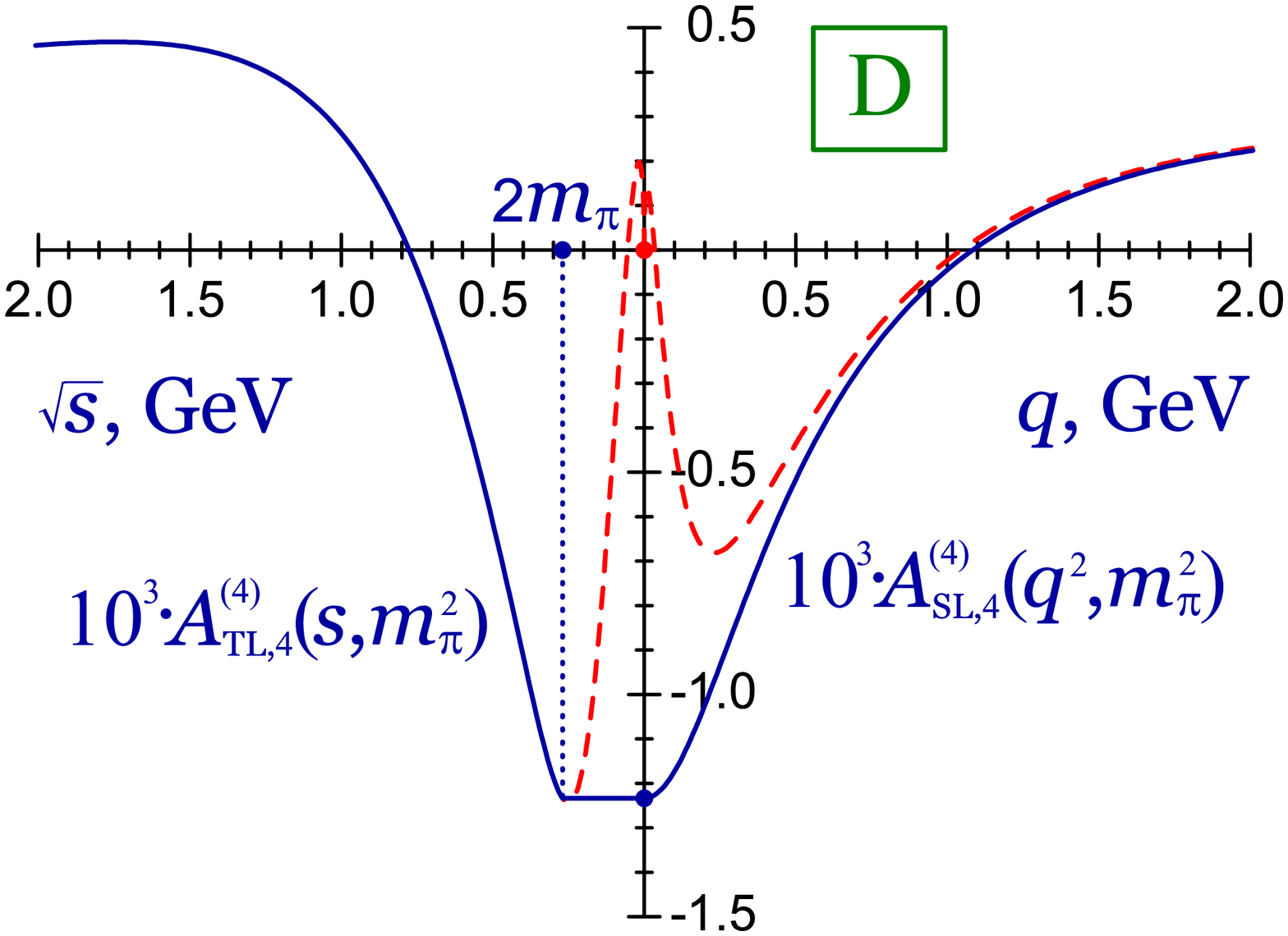}}\\
\end{tabular}}
\vskip-5.5mm
\caption{The four-loop nonpower expansion functions in spacelike
[$\ASL{(4)}{j}(q^2,\mpi^2)$, $q^2>0$, Eq.~(\protect\ref{ASLM})] and
timelike [$\ATL{(4)}{j}(s,\mpi^2)$, $s=-q^2>0$,
Eq.~(\protect\ref{ATLM})] regions. The plots A, B, C, and D
correspond to the first-, second-, third-, and fourth-order expansion
functions ($j=1,2,3,4$), respectively. The case of the nonvanishing
mass of the $\pi$~meson is denoted by solid curves, whereas the
dashed curves show the limit of the massless pion [see
Eqs.~(\protect\ref{ASLAPT})  and~(\protect\ref{ATLAPT})]. The values
of parameters are: $\Lambda=500\,$MeV, $\nf=2$.\vskip-0.5mm}
\label{Plot:NPEM4L}
\end{figure*}

     In particular, in the framework of the so--called analytic
perturbation theory (APT)~\cite{APT1,APT2,DV04} the pion mass  was
ignored in dispersion relation~(\ref{AdlerDisp}), and the analyticity
requirement of the form~(\ref{DefAn}) has been imposed on the
perturbative approximation~(\ref{AdlerPert}). Since both real and
imaginary parts of the strong running coupling contribute to the
relevant spectral density, eventually this led to the nonpower
expansion for the Adler function:
\begin{equation}
\label{AdlerAPT}
D(q^2) = 1 + \sum\nolimits_{j=1}^{\ell}d_{j} \ASL{(\ell)}{j}(q^2)
\end{equation}
(subscript ``SL'' stands for ``spacelike''), where
\begin{equation}
\label{ASLAPT}
\ASL{(\ell)}{j}(q^2) = \int_{0}^{\infty}
\frac{\varrho\ind{(\ell)}{j}(\sigma)}{\sigma + q^2}\, d \sigma
\end{equation}
and
\begin{equation}
\label{RhoAPT}
\varrho\ind{(\ell)}{j}(\sigma) = \frac{1}{\pi}
\lim_{\varepsilon \to 0_{+}} \mbox{Im}\left\{\!
\left[a\ind{(\ell)}{s}(-\sigma - i\varepsilon)\right]^{j}\right\}.
\end{equation}
In turn, the continuation of $D(q^2)$ (\ref{AdlerAPT}) into timelike
domain~(\ref{AdlerInv}) can also be represented in a form of the
nonpower functional expansion:
\begin{equation}
\label{RepemAPT}
R(s) = 1 +
\sum\nolimits_{j=1}^{\ell}d_{j} \ATL{(\ell)}{j}(s)
\end{equation}
(subscript ``TL'' stands for ``timelike'' here), with 
\begin{equation}
\label{ATLAPT}
\ATL{(\ell)}{j}(s) = \int_{s}^{\infty}\!
\varrho\ind{(\ell)}{j}(\sigma)\,\frac{d \sigma}{\sigma}.
\end{equation}
Since the functions~(\ref{ATLAPT}) automatically take into account
the so-called $\pi^2$--terms, the expansion coefficients~$d_j$
in Eqs.~(\ref{AdlerAPT}) and~(\ref{RepemAPT}) are identical. The
first--order expansion functions ($j=1$) correspond to the
$\ell$-loop running couplings in spacelike~(\ref{ASLAPT}) and
timelike~(\ref{ATLAPT}) domains, whereas the higher--order functions
($2 \le j \le \ell$) play the role of their effective powers. The
sets of those functions form the $\ell$-loop bases of the nonpower
expansions. Remarkably, the spacelike  functions~(\ref{ASLAPT})
deviate from the perturbative expansion basis at rather high
energies. Thus,  $\ASL{(4)}{1}(q^2)$ differs by $20\,\%$ from
$a\ind{(4)}{s}(q^2)$ at $q \simeq 2\,$GeV, whereas the difference
between  $\ASL{(4)}{4}(q^2)$  and $[a\ind{(4)}{s}(q^2)]^4$ is
$40\,\%$ at $q \simeq 10\,$GeV (see also Refs.~\cite{APT1,APT2,DV04} for
the details).

     In fact, the effects due to the masses of the light hadrons can
be safely neglected only when one handles the strong interaction
processes at high energies. However, in the intermediate- and
low-energy regions such mass effects become substantial. So, for the
case of the nonvanishing pion mass, one can bring the perturbative
expansion~(\ref{AdlerPert}) in conformity with the dispersion
relation~(\ref{AdlerDisp}) by requiring the former to satisfy the
integral representation of the form (see Ref.~\cite{MAIC})
\begin{equation}
\label{AdlerKLM}
D(q^2,\mpi^2) = \int_{4\mpi^2}^{\infty}
\frac{\varkappa(\sigma)}{\sigma + q^2}\, d\sigma.
\end{equation}
Ultimately, this also leads to the nonpower expansions for the Adler
function and $R_{e^{+}e^{-}}$~ratio:
\begin{equation}
\label{AdlerNPEM}
D(q^2,\mpi^2) = \frac{q^2}{q^2+4\mpi^2} \!+\!\!
\sum_{j=1}^{\ell}d_{j}\ASL{(\ell)}{j}(q^2,\mpi^2),
\end{equation}
where
\begin{equation}
\label{ASLM}
\ASL{(\ell)}{j}(q^2,\mpi^2) = \int_{4\mpi^2}^{\infty}
\frac{\varrho\ind{(\ell)}{j}(\sigma)}{\sigma + q^2}\, d \sigma
\end{equation}
[the spectral function~(\ref{RhoAPT}) is adopted herein]. It is worth
noting that $\ASL{(\ell)}{1}(q^2,\mpi^2)$ is a process--dependent
quantity, which can be identified with the QCD invariant charge at
high energies only, where the influence of the pion mass on
Eq.~(\ref{ASLM}) is negligible. Then, the  continuation
of~$D(q^2,\mpi^2)$ [see Eq.~(\ref{AdlerNPEM})] into timelike
domain~(\ref{AdlerInv}) reads
\begin{equation}
\label{RepemNPEM}
R(s,\mpi^2)=\theta(s-4\mpi^2)\!+\!\!
\sum_{j=1}^{\ell}d_{j} \ATL{(\ell)}{j}(s,\mpi^2),
\end{equation}
where $\theta(x)$ is the Heaviside step function and
\begin{equation}
\label{ATLM}
\ATL{(\ell)}{j}(s,\mpi^2) = \int_{s}^{\infty}\!
\theta(\sigma - 4\mpi^2)\,
\varrho\ind{(\ell)}{j}(\sigma)\frac{d \sigma}{\sigma}.
\end{equation}
It is worth emphasizing here that the main impact of the mass of the
$\pi$~meson on Eqs.~(\ref{AdlerNPEM}) and~(\ref{RepemNPEM}) is
twofold; not only the strong corrections, but also the parton model
predictions are modified at low energies (see  Ref.~\cite{AdlerIR}
for the details).

     The four-loop massive nonpower expansion functions~(\ref{ASLM})
and~(\ref{ATLM}) are presented in Figure~\ref{Plot:NPEM4L}. It turns
out that all the appealing features of the massless
APT~\cite{APT1,APT2,DV04} persist in the considered case of the
nonvanishing pion mass. Specifically, the functions~(\ref{ASLM})
and~(\ref{ATLM}) have no unphysical singularities and contain no
additional parameters. The timelike expansion functions~(\ref{ATLM})
automatically take into account the $\pi^2$--terms. The higher--order
functions are suppressed with respect to the preceding ones, a fact
that ultimately leads to the higher loop and scheme stability of
outcoming results.  It is worth noting that the spacelike expansion
functions~(\ref{ASLM}) are influenced by the pion mass in the
infrared domain below few GeV, whereas the timelike expansion
functions~(\ref{ATLM}) are affected by the mass of the $\pi$~meson
only below the two--pion threshold, where they acquire characteristic
plateaulike behavior (see Figure~\ref{Plot:NPEM4L}).

     The impact of the effects due to the nonvanishing mass of the
$\pi$~meson on the model~(\ref{AIC}) (see Ref.~\cite{AIC}) and  on
the processing the experimental data on the inclusive $\tau$~lepton
decay has been  discussed in detail in Ref.~\cite{MAIC}.

\section*{Acknowledgments}

\noindent
Authors wish to thank D.~Shirkov, K.~Chetyrkin, A.~Dorokhov,
S.~Kluth, S.~Narison, P.~Raczka, I.~Solovtsov, and V.~Zakharov for
the stimulating discussions and useful comments, as well as the
organizers of QCD$\:$05 for their hospitality. The work has been
supported by grants SB2003-0065 of the Spanish Ministry of Education,
CICYT FPA20002-00612, RFBR 05-01-00992 and NS-2339.2003.2.

\end{document}